# Increase of $YBa_2Cu_3O_7$ critical currents by Mo substitution and high-pressure oxygen annealing

A. Los [1], B. Dabrowski [2,*], and K. Rogacki [1]

[1] *Institute of Low Temperature and Structure Research, Polish Academy of Sciences, Okolna 2, 50-422 Wroclaw, Poland*

[2] *Institute of Physics, Polish Academy of Sciences, Aleja Lotnikow 32/46, 02-668 Warsaw, Poland*

[*] *Previous address: Department of Physics, Northern Illinois University, De Kalb, IL 60115, USA*

e-mail: *k.rogacki@intibs.pl*

We have previously shown for powders that Mo substitution into the CuO chains of $YBa_2Cu_3O_7$ can create effective pinning centres which significantly increase the critical current density ($j_c$) in 7 T field by a factor of 4 and 10 at 50 and 60 K, respectively [1]. The present work reports on the influence of the Mo substitution and high-pressure oxygen annealing on the pinning properties and critical currents of $YBa_2Cu_{3-x}Mo_xO_{7-d}$ by comparing pure (x = 0, d > 0) and substituted (x = 0.03, d < 0) single crystals. Pinning properties have been investigated by measurements of magnetization loops and calculations of $j_c$ in the *ab*-plane, in the temperature range from 2 to 90 K and in fields up to 14 T. Depending on the Mo substitution and the oxygen treatment, several types of pinning centres increasing $j_c$ have been revealed and analysed in the frame of Dew-Hughes' and Kramer's models.

1. Introduction

Ceramics of $YBa_2Cu_3O_7$ (YBCO) are known as the best superconductors for practical large-scale power applications at liquid nitrogen temperatures where there is no need for expensive cryogenic cooling devices. Achieving wider use of YBCO requires improving superconducting transport properties, including the critical current density ($j_c$) in high magnetic fields. For this reason, it is crucial to investigate vortex dynamics in a deep mixed state to find a way to enhance the vortex pinning force. Accordingly, as YBCO is characterized by relatively high anisotropy, study of the textured samples or single crystals is indispensable.

Increasing $j_c$ relays on enhancing of the flux pinning force which depends on the interaction of magnetic vortices with crystal defects. For this reason, the nano and micro sized impurities [2,3,4,5] or irradiation generated defects [6,7,8,9,10] are used as pinning centres. Depending on the type of the pinning centres, $j_c$ may be increased at different ranges of fields and temperatures. For example, for YBCO in the wide range of temperatures (20 - 85 K), the addition of microscopic grains of the $Y_2BaCuO_5$ phase works effectively in higher fields [2,5] while columnar defects created by neutron irradiation increase $j_c$ in lower fields [6]. However, all these methods are not easy implemented on a large scale, so simpler methods are desired. One of the most promising approaches is the optimization/modification of the YBCO structure by relatively simple chemical methods, which we study in this work. In this regard, several attempts have been made with varying degrees of success. For example, several transition elements (Al, Fe, Co, Ni, Zn) have been substituted for Cu in $CuO_2$ planes and/or CuO chains [11,12,13,14]. Since the planes are critically responsible for superconductivity, the substitutions in this region were ineffective as the critical temperatures ($T_c$) decreased rapidly from 92 K [11,12]. For less destructive substitutions for Ba with Sr [4,15] and large rare earths [16,17,18,19], no substantial increase of $j_c$ has been observed and the $T_c$'s also decreased. Rare earth ions have been as well substituted for Y [20,21,22,23,24,25] with some success as the $T_c$'s increased to 96 K for Nd and some increase of $j_c$ was achieved [22]. Another way to alleviate destruction of the $CuO_2$ planes superconductivity is the selective formation of changes (defects) in the CuO chains [11,26]. By applying this idea, we have previously used the Mo substitution and the high-pressure oxygen annealing in polycrystalline samples to create effective pinning centres in form of nano-sized defects in the CuO chain region of the YBCO structure [1,4].

The effectiveness of the Mo-based pinning centres has been shown in our studies performed on powdered samples [1]. It was discovered that after optimized synthesis in oxygen near 900 °C, the Mo substitutes for Cu in the CuO chains and creates dimers of the corner shared $MoO_6$ octahedra, $Mo_2O_{11}$, pointing preferentially in the a-axis direction, thus reducing the orthorhombic distortion (see Fig. 1). These defects after high-pressure oxygen anneal below 650 °C bring 3 extra oxygens per 2 $Mo^{6+}$ ions, which slightly increases charge doping. The $Mo_2O_{11}$ dimmers work as efficient pinning centres especially at low temperatures, where the coherence length in the *ab*-plane (~ 1.2 nm, [13,27]) is comparable with the size of the dimer (~ 0.4 nm x 0.8 nm). Moreover, an individual $Mo_2O_{11}$ dimer creates a crystal lattice distortion extending to 3-4 (or more) unit cells, which corresponds to the coherence length at higher temperatures, close to the boiling point of nitrogen, and therefore, the lattice distortion could work as a satisfactory pinning centre also at those temperatures.

Most applications of YBCO require epitaxially grown thin films due to relatively large anisotropy of this material [28,29,30]. Unfortunately, even for epitaxially grown films, the twin or grain boundaries are present, which mask the real nature of pinning acting on vortices in the volume. For this reason, studies using single crystals with well-defined orientation of their crystallographic directions relative to the direction of the applied magnetic field are desired to conclusively determine the flux pinning properties. Consequently, the pinning mechanism can be analysed for the well-defined field-current configuration, which for the most large-scale power applications means the direction of the magnetic field perpendicular to the current flowing in the *ab*-plane of a superconductor. In this work, we study the properties of the pinning force and the critical current density of high-quality YBCO single crystals modified by substituting Mo into the CuO chains and/or introducing additional oxygen, both of which create desirable pinning centres.

2. Experiment and samples characterization

Single crystals of $YBa_2Cu_{3-x}Mo_xO_{7-d}$ with nominal content $0 \leq x \leq 0.12$ were obtained by self-flux growth method from mixtures of single phase $YBa_2Cu_3O_{6.94}$ with $BaCuO_2$, $MoO_3$ and CuO. Mixtures were pressed into large-sized pellets (~ 10 g each) and pre-fired at 900 °C in oxygen. Sintered pellets of a total weight of about 40 g were loaded into Y stabilized $BaZrO_3$ crucibles and fired in air at 1040 °C for 15 hours followed by very slow cooling to 850 °C at 1.2 deg/hour and fast cooling to room temperature. The resulting melted mass contained in the crucibles was broken exposing internal cavities containing small, frequently self-staining plane-like crystals. Notably, attempts to grow these crystals in oxygen were not successful.

Normal pressure anneals were done in pure oxygen for 12 h at 300-500 °C followed by slow cooling to room temperature. Low and high pressure anneals were done in home-build apparatus in pure oxygen using 130-140 bar (LPA) at 250-400 °C and 255 bar (NS1) or 270 bar (HPA) at 500 °C, followed by slow cooling to room temperature at 6 deg/hour.

Several batches of pure and Mo-substituted YBCO single crystals were examined to select samples for further studies. The selection was performed in two stages. The first stage was based on the Energy Dispersive X-ray Spectrometry (EDXS) investigations, which revealed a proper ratio of the elements Y, Ba and Cu, and their uniform distribution, except for Mo, which showed significant dispersion for some batches, as presented in Fig. 2. Single crystals from batch named YMo2, with the Mo nominal content of $x = 0.03$ (1 at. %), which showed the most uniform Mo EDXS content of x ~ 0.015 (0.5 at. %), have been chosen for further studies. This corresponds to 1.5 at. % substitution of Mo in the CuO chains, and to about 1/133 planar density of the $Mo_2O_{11}$ dimmers.

In the next stage of crystals selection the magnetization measurements were used to determine the superconducting properties of the samples. Single crystals with the highest critical temperature and the sharpest superconducting transition, which indicated the high phase homogeneity, purity and optimal oxygen content, were selected for investigation of the critical currents. Examples of the transition to the superconducting state are shown in Fig. 3, for two pure (Y123, $x = 0$) and two Mo-substituted (YMo2, $x = 0.03$) single crystals. The selected samples were parallelepipeds with dimensions of about $1.1 \times 1.0 \times 0.20$ mm$^3$ for pure, and $1.0 \times 0.8 \times 0.15$ mm$^3$ for Mo-substituted samples, such that the dimensions had similar ratios.

Critical current densities were obtained from magnetization hysteresis loops by using the anisotropic Bean model, in which the critical currents flowing in the *ab*-plane have a density of $j_c^{ab} \approx 20\Delta M/b(1-b/3a)$ and in the *c*-axis direction of $j_c^c \approx 20\Delta M/b$, for $a \geq b$, where *a* and *b* are the sample dimensions in the *ab*-plane, and $\Delta M$ is the difference between the upper and the lower branches of the hysteresis loop. In this work, we report only the $j_c^{ab}$, i.e. the critical current density obtained for the magnetic field oriented parallel to the *c*-axis direction. This configuration, which is present for most applications excludes the influence of the strong pinning within the *ab*-plane, and therefore makes our pinning analysis simpler and more reliable. The magnetization loops have been measured in the temperature range of 2 to 90 K and in magnetic fields up to 7 and 14 T using the MPMS and PPMS magnetometers, respectively.

3. Results and discussion

A sequence of magnetization measurements was performed to differentiate between the extrinsic and intrinsic effects of the Mo-substitution and oxygen annealing on the observed enhancement of $j_c$, and to relate it to the flux pinning by $MoO_6$ dimers in the CuO chains. Magnetization versus temperature in the vicinity of the superconducting transition is shown in Fig. 3 for all single crystals selected for studies. For these measurements, the zero-field-cooling (ZFC) and (for sample YMo2-130b400-NS1) field-cooling (FC) procedures were used, and the magnetization was normalized to its value at $T = 10$ K. All samples show sharp superconducting transitions except for YMo2-130b400, which most likely had insufficient oxygen content, because the Mo-substituted single crystals could not be sufficiently oxygenated without additional annealing under high oxygen pressure. The FC values obtained for YMo2-130b400-NS1 are relatively small indicating strong pinning properties.

Examples of magnetization loops for a pure single crystal annealed in oxygen at low pressure are shown in Fig. 4, for the virgin (the beginning of the $M(H)$ curve at $M(H=0) = 0$) and mixed superconducting states. For the $M(H)$ loops, the second magnetization peak (SMP) occurs in higher fields (e.g., in ~7 T at 15 K) and shifts to lower fields as the temperature increases. Examples of $M(H)$ loops at temperatures closer to $T_c$ are shown in Fig. 5, for the Mo-substituted single crystal, annealed in oxygen, first at low and then at high pressure. The increase in the width of the loops, and therefore in $j_c$, due to the high pressure annealing is clearly visible. Furthermore, the position of the SMP shifts towards higher fields. Both effects indicate that annealing under high oxygen pressure increases the number or effectiveness of the pinning centres, as we discuss in the following paragraphs.

The observed SMP is not geometric in nature [31,32,33], because it remains practically unchanged after breaking the crystal into several pieces. This is consistent with the general understanding of SMP in YBCO, as discussed by Pasquini *et al.* [34]. The SMP is also not caused by the so-called matching effect, if we assume that the amount of flux quanta for given $B$ fits to the number of pinning centres defined as the individual $Mo_2O_{11}$ octahedra dimers. Considering the amount of substituted Mo, we calculate that the surface density of the dimers in the *ab*-plane is ~27·10$^{16}$ m$^{-2}$ (assuming even volume distribution of the unit cells containing dimers) or ~14·10$^{16}$ m$^{-2}$ (assuming even distance between dimers), which corresponds to $B_{mat}$ ~ 560 or 290 T, respectively. Therefore, in order to explain the behaviour of SMP in our single crystals, the mechanisms of clustering of individual pinning centres and/or collective pinning have been assumed, as considered in [35] and [1], respectively.

Magnetization hysteresis loops similar to those exemplified in Figs. 4 and 5 were used to calculate the density of critical currents for our samples. The characteristic features of the $M(H)$ loops relate to the properties of $j_c(H) \sim \Delta M(H)$, which we will discuss now. Critical currents of the pure Y123 and Mo-substituted YMo2 single crystals annealed in oxygen under low pressure (LPA) are shown in Fig. 6. For YMo2, at low temperatures (4 K), a slight increase in $j_c^{ab}$ is observed in the whole range of the applied fields (0-9 T). At higher temperatures, the increase in $j_c^{ab}$ occurs for fields below 2.5 T. This increase is due to a peak effect (PE) which appears for the $j_c^{ab}(H)$ dependence as a result of additional pinning centres created by the Mo substitution. The additional pinning centres are most likely $Mo_6O_{11}$ dimers, the occurrence of which has been shown for Mo-substituted YBCO ceramics by the neutron powder diffraction, the oxygen thermogravimetry and analysis of pinning properties [1,4]. The decrease of $j_c^{ab}$, observed for $B > 2.5$ T or $T > 70$ K, for the Mo-substituted single crystal, seems to be caused by the worsening of the superconducting properties, as was shown in Fig. 3.

Critical currents of Y123 and YMo2 single crystals annealed in oxygen under high pressure (HPA and NS1) are shown in Fig. 7. At low temperatures, no large difference has been observed for pure and Mo-substituted crystals, however at temperatures above 50 K, a clear increase of $j_c^{ab}$ has been revealed for the YMo2-NS1 sample, in the whole range of the applied magnetic fields. This increase is particularly significant in low fields, where in a 2 T field it is 20 and 100 times at temperatures of 70 and 75 K, respectively. As in the case of Mo-substituted LPA single crystal, the increase of $j_c^{ab}$ can be explained by the appearance of PE, which for Mo-substituted YMo2-NS1 single crystals is present at higher fields, e.g., 4 and 1 T at 55 and 80 K, respectively. The expected shift of PE to lower fields with increasing temperature is expected to be caused by the increase in the coherence length, which results in the reduction of the effective number of pinning centres, as now one vortex core contains several such centres.

Useful conclusions can also be drawn by comparing the $j_c$ for single crystals annealed in oxygen at low (Fig. 6) and high (Fig. 7) pressure. For pure Y123 crystals annealed at high pressure, $j_c$ is larger at temperatures up to 60 K, however clearly smaller at higher temperatures becoming not measurable at 80 K. This behaviour seems to be caused by the appearance of interstitial oxygen, which increases the number of point pinning centres effective at low temperatures (at smaller coherence length), but simultaneously reduces the number of more extended pinning centres, effective at higher temperatures (at larger coherence length), created by the clustered oxygen vacancies, as shown in [35]. The huge increase of $j_c$ due to the annealing in oxygen at high pressure is observed for the Mo-substituted single crystal YMo2 (Fig. 7(b)). This increase of $j_c$ is present at all temperatures, at which the measurements were performed, and throughout the entire range of the applied magnetic fields (here up to 14 T). This is most likely due to the proper oxygenation of the single crystal, which enables the formation of stoichiometric $Mo_6O_{11}$ octahedra dimers that (whenever clustered or not) deform locally the crystal structure. Additional results will be drawn by analysing the behaviour of the pinning force, which will be presented in the next paragraphs.

For studying the nature of pinning effects, we used the Dew-Hughes scaling, which predicts the dependence of the volume pinning force, $F_p = \mu_0 H j_c$, normalized to its maximum value, $F_p^{max}$, on the reduced magnetic field, $h = H/H_{c2}$, where $H_{c2}$ is the upper critical field. The scaling formula $f_p = F_p/F_p^{max} = A \cdot h^p (1-h)^q$ is obtained, where A is the proportionality constant, and $p$ and $q$ characterize the type of pinning centres [36]. For high temperature superconductors, instead of $H_{c2}$ the irreversibility field, $H_{irr}$, is usually used, which we have determined with the Kramer's approach [37]. This method involves the linear extrapolation of the relation $y = j_c^{1/2} H^{1/4}$ (the so-called Kramer's plot) to $y = 0$ (so $j_c = 0$), which allows to determine the critical field $H_K = H(j_c=0) = H_{irr}$ for individual temperatures [1,38,39]. Because $H_K$ obtained this way is a lower limit of $H_{irr}$ when measured directly (e.g., in $M(H)$ or $R(T)$ experiments), it describes properties related to the global critical current flowing through the entire single crystals rather than the maximal critical current preserved in some local areas. One advantage of using $H_K$ is that it can be derived at fields much higher than those available in actual measurements. In our case, it was possible to determine $H_K$ for temperatures down to 35 K, where $H_{irr}$ is too large to be measured directly.

Figures 8 and 9 show the field dependences of the normalized pinning force, $f_p(h)$, where $h = H/H_K$, for all single crystals reported here. The Figures also show the model relations $f_p = A \cdot h^p (1-h)^q$ for different values of $p$ and $q$, to identify the type of dominant pinning centres operating in our samples [36]. For pure Y123 single crystals, annealed in oxygen both at low (Fig. 8(a)) and high (Fig. 9(a)) pressure, the scaling is rather poor and difficult to interpret. Moreover, no pinning centres are dominant when temperatures are changing from 70 to 80 K and from 45 to 75 K, for crystals annealed in low and high oxygen pressure, respectively. For crystal Y123-140b250, it was also not possible to obtain a clear criterion for determining $H_K$ below 70 K due to lack of a linear relationship for Kramer's plots. On the other hand, the Mo-substituted YMo2 single crystals show quite good scaling, especially for the single crystal annealed in oxygen at high pressure (NS1). Comparing the experimental results with the Dew-Hughes formula $f_p(h)$ we found that the best fit is observed with $p = 0.5$ and $q = 2$, for YMo2-130b400 (Fig. 8(b)). This indicates that normal (non-superconducting) surface-like pinning centres dominate in this sample, at least at higher fields, above the $f_p(h)$ maximum [36]. We interpret these pinning centres as the $Mo_6O_{11-d}$ dimers not fully oxygenated because the YMo2-130b400 was annealed in oxygen at insufficient pressure. These dimmers are rather smeared, weak defects and can only work as medium strength pinning centres. Annealing in oxygen at high pressure forces the full oxygenation of the dimers and, additionally, the creation of interstitial oxygen, both working as the well-defined point pinning centres. These pinning centres may also cluster forming larger defects, for example, volume-like four $MoO_6$ octahedra replacing a square of four Cu ions [1], as the clustering was considered when explaining the behaviour of the second magnetization peak. These estimates are consistent with the results obtained for YMo2-NS1 single crystal (Fig. 9(b)), as the best fit to the $f_p(h)$ curves was found with $p = 1$ and $q = 1$ at low fields and with $p = 1$ and $q = 2$ at higher fields, which indicates that the $\Delta\kappa$ volume-like and the normal point-like pinning centres dominate, respectively. The "normal" pinning centers may include areas with reduced $T_c$, while the $\Delta\kappa$ pinning centers dominate in areas with the altered Ginzburg-Landau parameter [36]. The type of interaction of all forms of pinning centres identified in our single crystals is "vortex core" in nature, because their sizes in all directions are smaller than the penetration depth of YBCO.

Strong pinning, observed for Mo-substituted single crystals, and a good scaling of the pinning force with the magnetic field in a wide temperature range from 35 to 80 K shows that one type of pinning centres is mainly responsible for the increase in critical currents that we observe. As we proved before for powders, the pinning centers are the $Mo_6O_{11}$ octahedra dimers, which for samples annealed in oxygen under high pressure may be supplemented by excess interstitial oxygen [1,4]. These pinning centers have a point-like character and are effective when the density of vortices is large for high fields. They may cluster forming pins with volume-like properties which are effective at lower fields (lower density of vortices), as observed. Results obtained for single crystals show that for the configuration when the magnetic field is perpendicular to the *ab*-plain, i.e., for the case when the $CuO_2$ planes are not important for pinning the vortices, it is possible to increase the critical currents by a factor of 2-5, depending on temperatures and fields, by substituting a small amount of Mo into the CuO chains.

4. Summary

Pure and Mo-substituted YBCO single crystals were synthesized in air and annealed in oxygen at elevated pressures of 130-140 bar (low) and 255-270 bar (high). Pinning properties have been analysed showing the efficiency of the $Mo_2O_{11}$ octahedra dimers and excess interstitial oxygen as pinning centres. For the Mo-substituted single crystals, the volume pinning force shows scaling in the wide range of temperatures, pointing on the $\Delta\kappa$ volume-like and the normal point-like pinning centres dominating at low and high magnetic fields, respectively. For pure YBCO, the annealing at high pressure increases the critical current density $j_c$ by a factor of 2-3, in a wide range of magnetic fields (0-9 T) at low temperatures (4-45 K). Substitution of Mo increases $j_c$ on a similar scale in the wider range of temperatures (4-80 K), but at lower fields. For single crystals substituted with Mo and annealed in oxygen at high pressure, the increase of $j_c$ by a factor of 3-5 was obtained, depending on temperatures and fields, which is desirable property for large-scale applications such as wires for power cables, generators and current transformers.

Acknowledgements: The research was partially supported by The National Centre for Research and Development (NCBR, Poland) within the project "ERA.Net RUS Plus: No146 – MAGNES" financed by the EU 7th FP, grant no 609556.

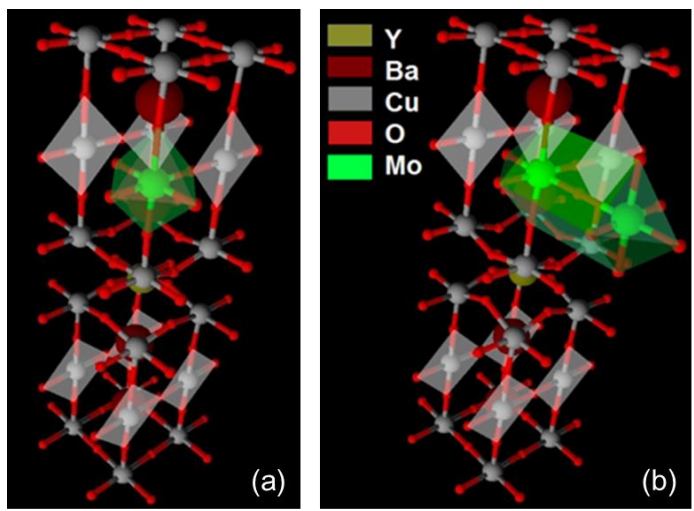

Fig. 1. Structure of the YBCO with CuO chains marked as a line of grey diamonds and the $MoO_6$ octahedra highlighted in green. a) Single $MoO_6$ octahedra bringing 2 extra oxygens per Mo ion and (b) $Mo_2O_{11}$ dimers pointing perpendicular to the CuO chain direction and bringing 3 extra oxygens per 2 Mo ions.

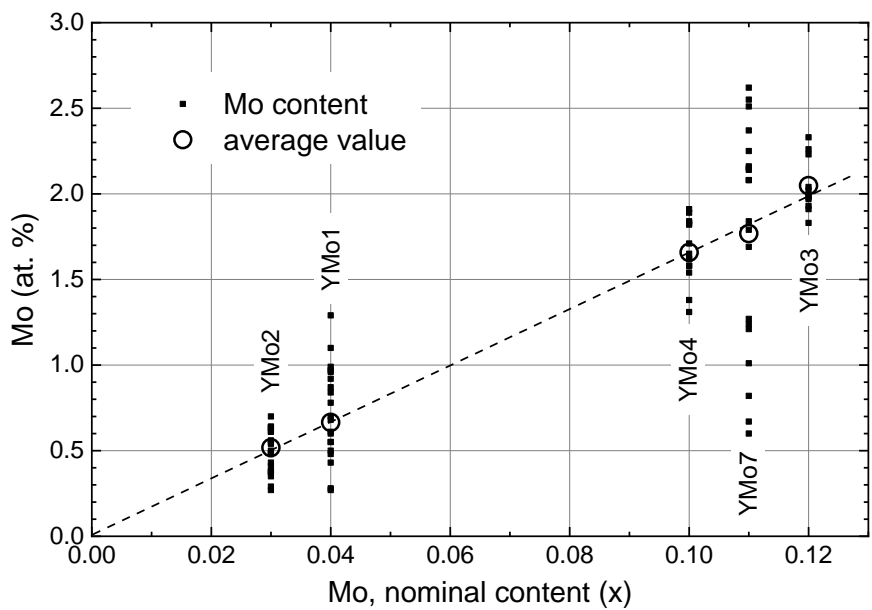

Fig. 2. Measured total atomic molybdenum content obtained by the EDXS versus the nominal Mo content, for several single crystals selected from five batches of $YBa_2Cu_{3-x}Mo_xO_{7-d}$. The average values (open circles) appear to be arranged along a straight line. Samples form the batch YMo2 have been investigated in this work.

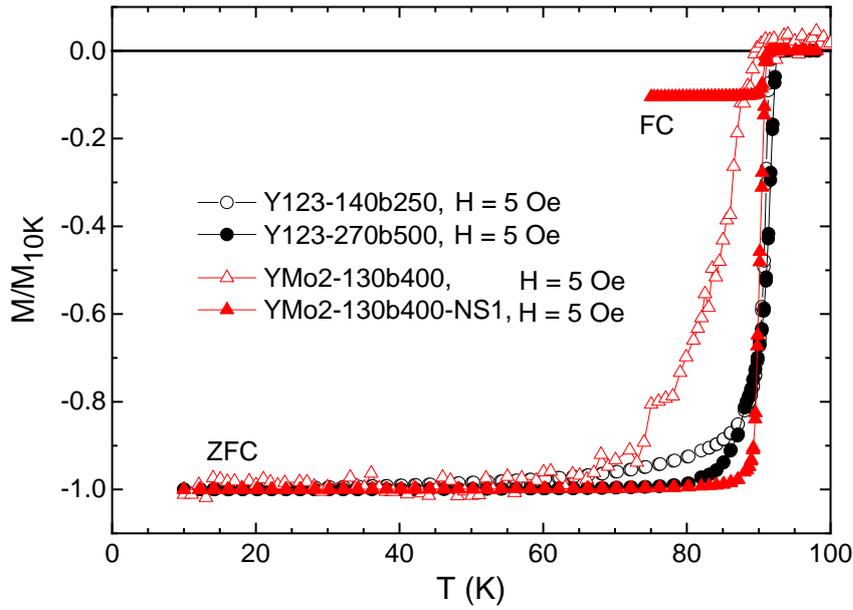

Fig. 3. Magnetization versus temperature for pure (Y123) and Mo-substituted (YMo2) YBCO single crystals annealed in oxygen under low (open symbols) and high (filled symbols) pressure. NS1 means the additional annealing at 255 bar and 500 °C of the same crystal. For the $M(T)$ measurements, zero-field-cooling (ZFC) and field-cooling (FC) procedures were used and the magnetization was normalized to its value at $T = 10$ K.

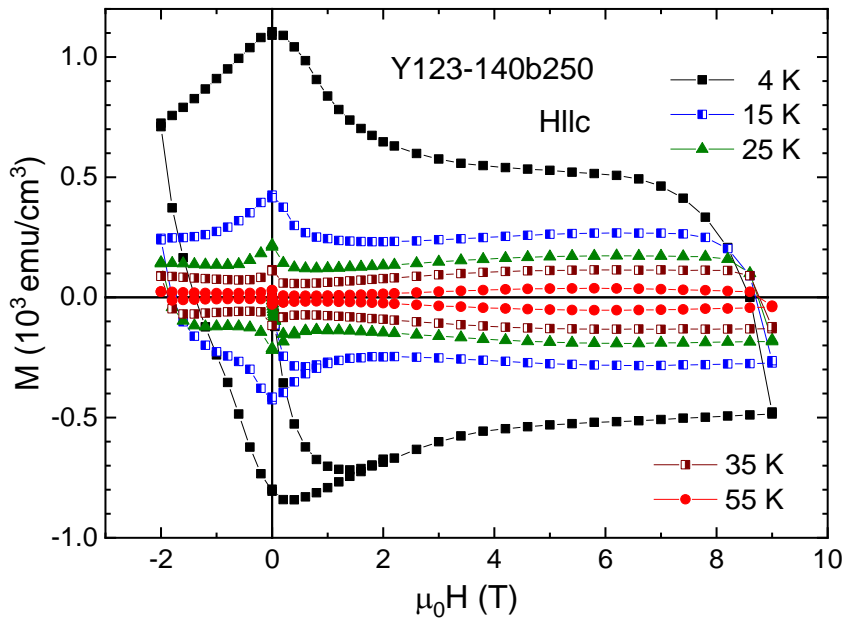

Fig. 4. Magnetization loops at several temperatures for the pure YBCO single crystal annealed in oxygen at low pressure (140 bar, 250 °C). The magnetic field is oriented parallel to the $c$-axis of the single crystals.

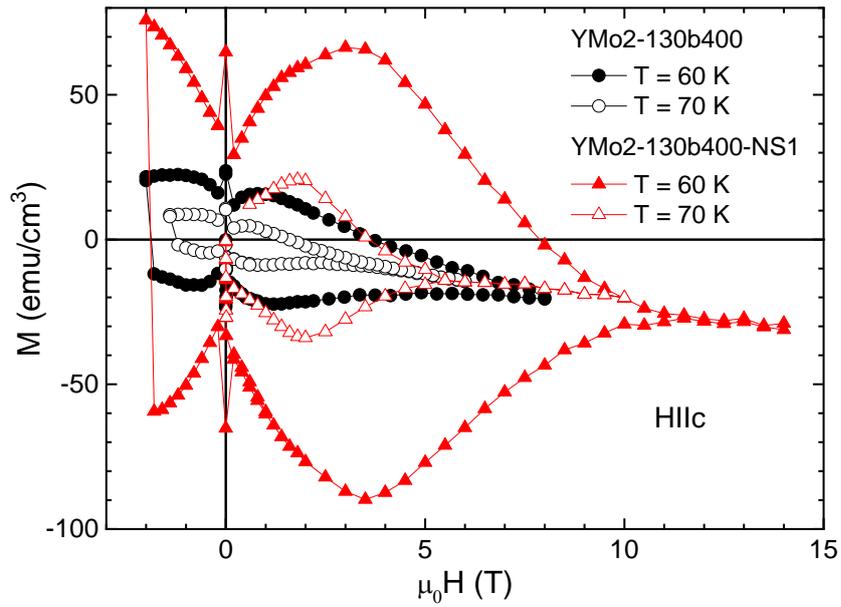

Fig. 5. Magnetization loops at 60 K (filled symbols) and 70 K (open symbols) for the Mo-substituted YBCO single crystal, annealed in oxygen at low pressure (YMo2-130b400, circles) and then additionally at high pressure (YMo2-130b400-NS1, triangles) at 255 bar and 500 °C. The magnetic field is oriented parallel to the *c*-axis of the single crystals.

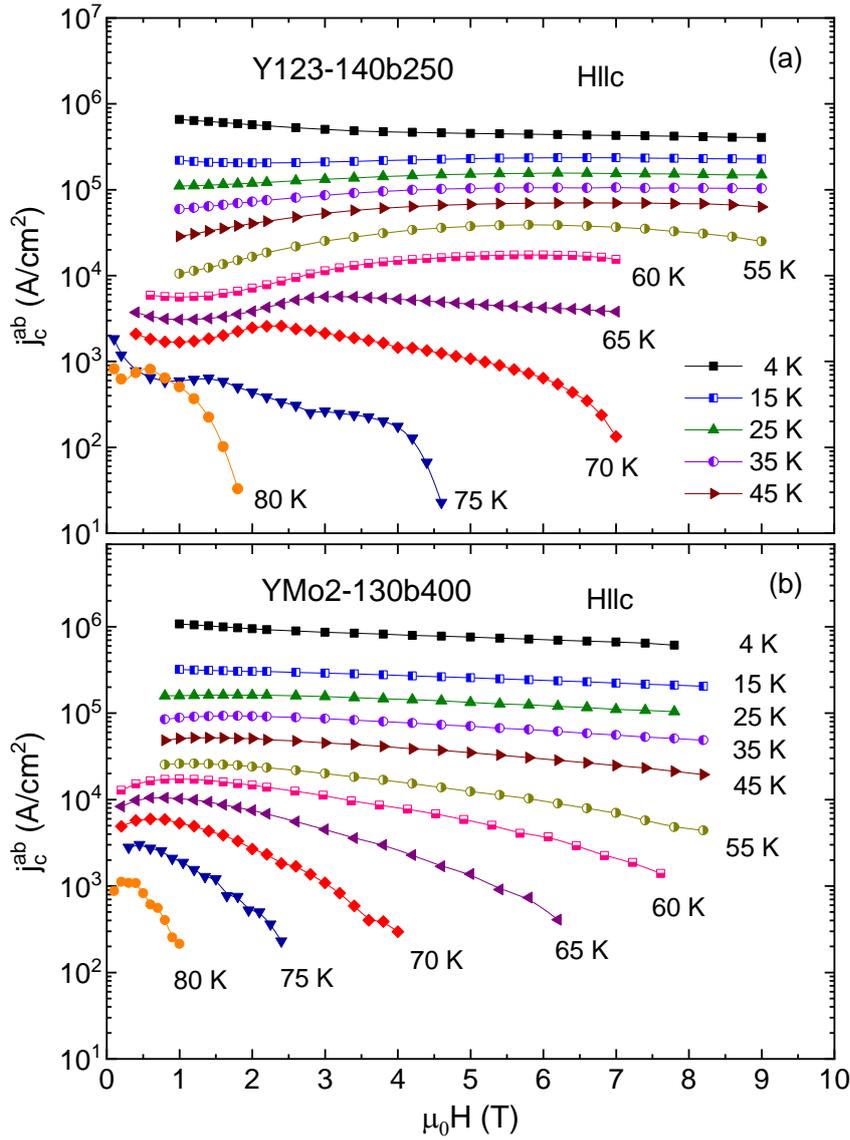

Fig. 6. Critical current densities vs magnetic field at constant temperatures for: (a) the pure YBCO single crystal (Y123-140b250) and (b) the Mo-substituted YBCO single crystal (YMo2-130b400), both annealed in oxygen at low pressure. The magnetic field is oriented parallel to the *c*-axis of the single crystals.

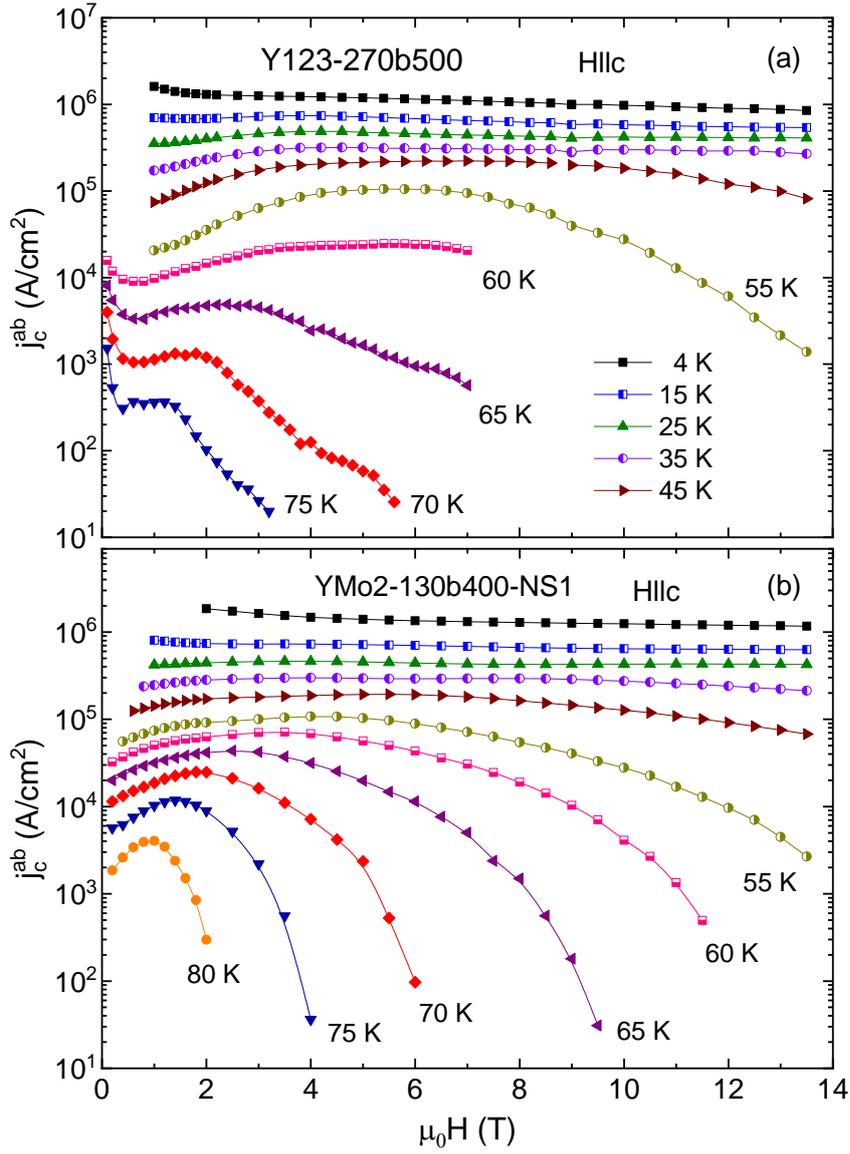

Fig. 7. Critical current densities vs magnetic field at constant temperatures for: (a) the pure YBCO single crystal annealed in oxygen at high pressure (Y123-270b500) and (b) the Mo-substituted YBCO single crystal annealed in oxygen at low and then additionally at high pressure (YMo2-130b400-NS1) at 255 bar and 500 °C. The magnetic field is oriented parallel to the *c*-axis of the single crystals.

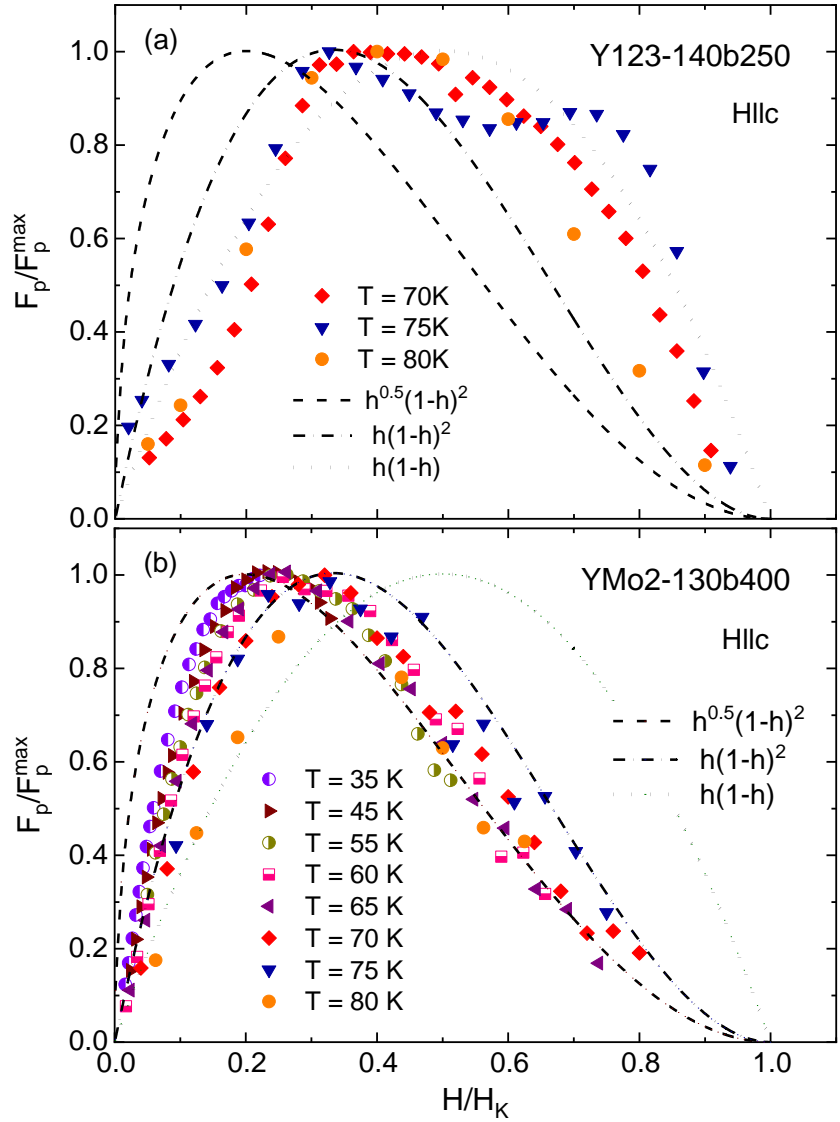

Fig. 8. Reduced pinning force versus reduced magnetic field at constant temperatures for: (a) the pure YBCO single crystal (Y123-140b250) and (b) the Mo-substituted YBCO single crystal (YMo2-130b400), both annealed in oxygen at low pressure. The lines show scaling curves $F_p/F_p^{max} = f_p = h^{0.5}(1-h)^2$ (dashed line), $f_p = h(1-h)^2$ (dashed dot line), and $f_p = h(1-h)$ (dot line), according to the Dew-Hughes scaling model, where $h = H/H_K$ and $H_K$ is the Kramer's scaling field.

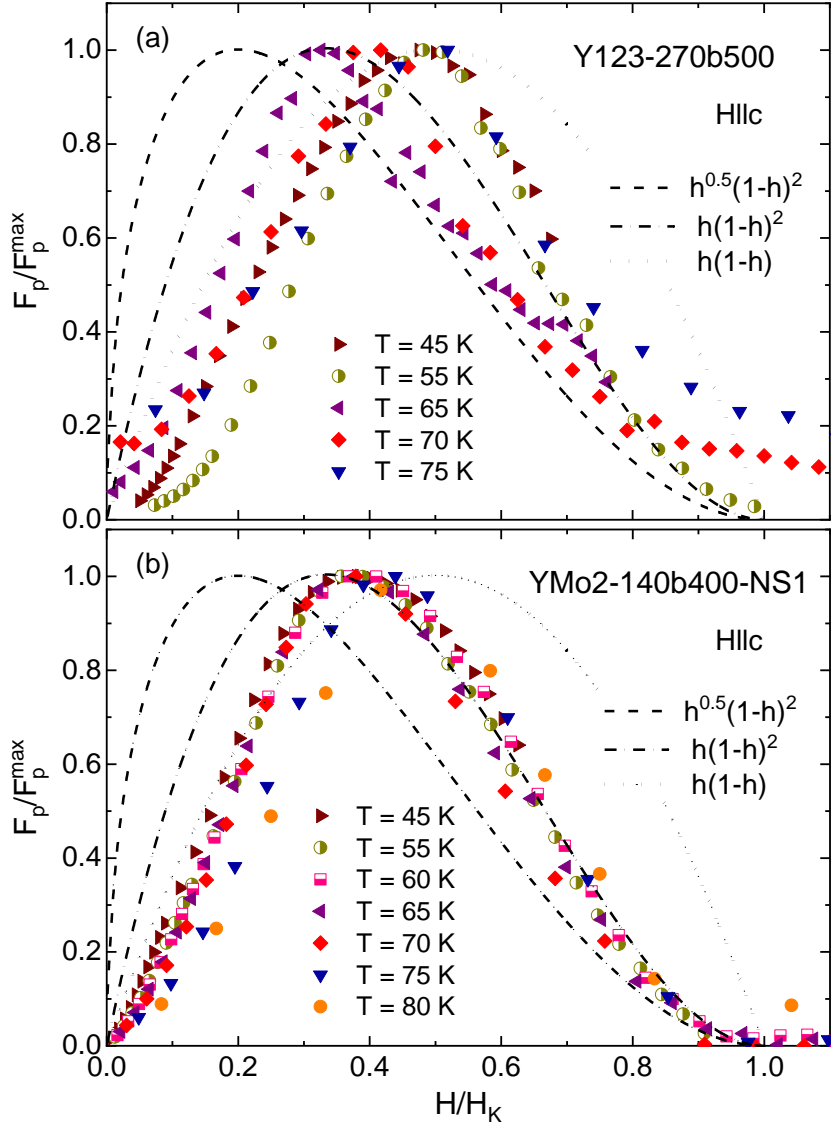

Fig. 9. Reduced pinning force versus reduced magnetic field at constant temperatures for: (a) the pure YBCO single crystal annealed in oxygen at high pressure (Y123-270b500) and (b) the Mo-substituted YBCO single crystal annealed in oxygen at low and then additionally at high pressure (YMo2-130b400-NS1) at 255 bar and 500 °C. The lines show scaling curves $F_p/F_p^{max} = f_p = h^{0.5}(1-h)^2$ (dashed line), $f_p = h(1-h)^2$ (dashed dot line), and $f_p = h(1-h)$ (dot line), according to the Dew-Hughes scaling model, where $h = H/H_K$ and $H_K$ is the Kramer's scaling field.